\documentclass[a4paper]{article}
\usepackage{a4}
\usepackage{fullpage}
\usepackage{amsmath,amssymb}
\usepackage{graphicx}
\usepackage{hyperref}
\usepackage{flafter}

\DeclareMathOperator{\Bin}{Bin}
\author{Yuri Yu. Tarasevich, Andrei V. Eserkepov, Irina V. Vodolazskaya}
\title{Electrical conductivity of randomly placed linear wires: a mean field approach}

\begin{document}
\maketitle

\begin{abstract}
  Using the mean-field approximation, a formula for the effective electrical conductivity of a two-dimensional system of randomly arranged conducting sticks with a given orientation distribution was obtained. Both the resistance of the sticks themselves and the resistance of the contacts between them were taken into account. The accuracy in the resulting formula was analyzed. A comparison of the theoretical predictions of mean-field approach with the results of direct electrical conductivity calculations for several model orientation distributions describing systems with crossed sticks demonstrated good agreement. Our study showed that cross-alignment of nanowires should lead to a decrease in the electrical conductivity compared to electrodes with isotropically arranged nanowires. We suppose that the widely used model with zero-width sticks is quite acceptable for systems of cross-aligned nanowires, but overestimates their connectivity in isotropic systems. Thus, the enhancement of the electrical conductivity of conducting films with cross-aligned nanowires may be due to a significant difference in the network topology.
\end{abstract}

\section{Introduction}\label{sec:intro}
Modern technologies make it possible to produce conductive films consisting of two layers of nanowires: in the first layer, the nanowires are aligned predominantly along one direction, while in the second layer, they are aligned perpendicularly. Examples of such technologies are dip coating~\cite{Duan2015}, the capillary printing technique~\cite{Kang2015}, blade-coating method~\cite{Fang2016}, bar-coating assembly~\cite{Cho2017}, direct-writing solution process guided by a conical fiber array~\cite{Meng2021}, deposition by using the shear force from the high-speed rotating concentrated nanowire solution~\cite{Hu2023} or from Mayer's rod~\cite{Wang2024,Wang2025,Wang2025a}, a continuous roll-to-roll process~\cite{Yang2024}. The optoelectrical performance of films with a cross-aligned arrangement of metallic nanowires is superior to that of those with a random arrangement~\cite{Grazioli2025}.  The cross-aligned nanowires are favorable for their nanowelding, which plays a crucial role in reducing junction resistance between nanowires in electrodes based on metallic nanowires~\cite{Wang2025,Wang2025a}. Conductive films based on cross-aligned nanowires are more uniform compared to films based on randomly arranged nanowires, which reduces the likelihood of electrical breakdown and the appearance of hot spots (overheated areas)~\cite{Cho2017}. The surface roughness of these films is less as compared to films based on randomly arranged nanowires~\cite{Duan2015,Cho2017}. Experiments show that for the same transparency, which is determined by the concentration of nanowires, systems with cross-aligned nanowires have lower sheet resistance compared to isotropic systems~\cite{Fang2016,Cho2017}.

In the case of random nanowire networks, the quasi-three-dimensional (Q3D) topology of the real network may significantly differ from the two-dimensional (2D) topology used in computer simulations when nanowires are modeled by zero-width sticks~\cite{Daniels2021}. Contrary, in the case of networks of welded cross-aligned metallic nanowires, the difference between real-world networks and their 2D models are expected to be insignificant.

Recently, the percolation threshold, electrical conductivity and homogeneity of nanowire networks having cross-aligned and random arrangements have been extensively investigated by means of computer simulations~\cite{Grazioli2025}. The authors demonstrated that the superior optoelectronic properties of cross-aligned nanowire networks arise from the combined effect of a reduced junction resistance and improved deposition uniformity.

Previously, the effective electrical conductivity of random nanowire networks was obtained using the mean-field approximation (MFA) under the assumption that the contacts on the conductors are uniformly distributed, while all stick orientations are equiprobable~\cite{Tarasevich2022}. As a matter of fact, both these assumptions are unnecessary. In the present work,  within the framework of the MFA, we offer a more common and rigorous derivation of the effective electrical conductivity of nanowire networks when nanowires obey a given orientational distribution. Our main focus is networks with cross-aligned nanowires.

The rest of the paper is constructed as follows. Section~\ref{sec:methods} describes our model (Sec.~\ref{subsec:model}, characterises orientational distributions used in our study (Sec.~\ref{subsec:ODFs}), and presents some technical details of the simulation (Sec.~\ref{subsec:simul}). In Section~\ref{sec:results}, we present the improved formula for the effective electrical conductivity and discuss its accuracy (Sec.~\ref{subsec:analytics}), compare the result of the direct computations of the effective electrical conductivity with predictions of the MFA  (Sec.~\ref{subsec:compar}). Section~\ref{sec:concl} summarizes the main results. Mathematical details are presented in Appendix~\ref{sec:deriv}.

\section{Methods}\label{sec:methods}
\subsection{Model}\label{subsec:model}
Let $N$ identical zero-width conductive sticks of length $l$ be placed within a square domain $\mathcal{D}$ of size $L \times L$ ($L>l$) with periodic boundary conditions (PBCs).  The number density of sticks, i.e., the number of sticks per unit area, is
\begin{equation}\label{eq:numdens}
n = \frac{N}{ L^2}.
\end{equation}
The centres of the sticks are assumed to be independent and identically distributed (i.i.d.) within  $\mathcal{D} \in \mathbb{R}^2$, i.e., $x,y \in [0;L]$, where $(x,y)$ are the coordinates of the centre of the stick under consideration. Their orientations are assumed to obey a given orientational distribution function (ODF). In such a way, a homogeneous network is produced. Let the electrical resistance of each stick be $R_\text{s}$, while the resistance of each contact (junction) between any sticks be $R_\text{j}$.

The main idea behind a MFA is as follows. Instead of considering the entire ensemble of conductors, one can consider a single conductor located in the average electrostatic field created by all the other conductors. Since the derivation of the formulas is, with two exceptions, similar to that given in~\cite{Tarasevich2022}, we present in Section~\ref{subsec:analytics} only the final master equation; the interested reader can find detailed derivations in Appendix~\ref{sec:deriv}.

\subsection{Orientational distributions}\label{subsec:ODFs}
In our study several ODFs were used. For the isotropic case, the ODF is as follows
\begin{equation}\label{eq:ODF1}
  f_{\alpha}(\alpha) = \frac{1}{\pi},
\end{equation}
where $\alpha$ is the angle between the stick and the $x$ axis.
For this particular case, the computed electrical conductivity was compared with the MFA estimate in Ref.~\cite{Tarasevich2022}.

To simulate systems with cross-aligned nanowires, two normal probability distributions with standard deviation $\sigma  = \pi /11.5$ and mean values $\mu_1$ and $\mu_2$ such that $\mu_2 = \mu_1 +\pi /2$ were used in Ref.~\cite{Grazioli2025}
\begin{equation}\label{eq:ODFGauss}
  f_{\alpha}(\alpha) = \frac{1}{2\sigma\sqrt{2\pi}}\left[ \exp\left(-\frac{(\alpha-\mu_1)^2}{2\sigma^2}\right) + \exp\left(-\frac{(\alpha-\mu_2)^2}{2\sigma^2}\right)\right].
\end{equation}

Let the fraction $\omega$ of the sticks be oriented along the $x$-axis, while the fraction $1-\omega$ be oriented along the $y$-axis. In this case, the ODF is as follows
\begin{equation}\label{eq:ODF2}
f_{\alpha}(\alpha) = \omega \delta(\alpha) + (1-\omega)\delta\left(\frac{\pi}{2}-\alpha\right).
\end{equation}

When the orientations of sticks are equiprobable within two intervals of width $2\varepsilon$, the ODF is as follows
\begin{equation}\label{eq:ODF3}
f_{\alpha}(\alpha) =
\begin{cases}
  \dfrac{1}{4\varepsilon}, & \text{if } \alpha \in \left[-\dfrac{\pi}{2}; -\dfrac{\pi}{2} + \varepsilon\right] \cup [-\varepsilon;\varepsilon] \cup \left[\dfrac{\pi}{2}-\varepsilon; \dfrac{\pi}{2}\right],  \\
  \\
  0, & \mbox{otherwise}.
\end{cases}
\end{equation}

To obtain a smooth distribution with two humps, we used the following ODF
\begin{equation}\label{eq:ODF4}
f\left(\alpha ,m \right)=\frac{\Gamma  \left(\frac{m}{2}+1\right) }{\Gamma  \left(\frac{m}{2}+\frac{1}{2}\right) \sqrt{\pi}}{| \cos  \left(2 \alpha \right)|}^{m}.
\end{equation}
As a matter of fact, the ODF~\eqref{eq:ODF4} is a minor modification of the distribution that was used previously to simulate systems of elongated particles oriented predominantly along one direction~\cite{Chatterjee2014,Hoerrmann2014,Chatterjee2015,Klatt2017}.

Since for all above distributions
\begin{equation}\label{eq:cos2ODFs}
\left\langle \cos^2 \alpha \right\rangle = \frac{1}{2},
\end{equation}
the nematic order parameter is zero and therefore uninformative. However, the variance of the nematic order parameter~\cite{Chatterjee2014}
\begin{equation}\label{eq:S2}
  \left \langle S^2 \right\rangle = \int\limits_{-\pi/2}^{\pi/2} f(\alpha) \cos^2 \left(2\alpha\right) \, \mathrm{d}\alpha
\end{equation}
is quite convenient for characterizing the distributions.
Therefore, $\left \langle S^2 \right\rangle$ increases from 1/2 for equiprobable orientations of sticks to 1 for a perfect cross-alignment.

Quantities of interest for different ODFs are presented in Table~\ref{tab:ODFs},
where
\begin{equation}\label{eq:sin12}
\langle |\sin(\alpha_1 - \alpha_2)| \rangle =
\int\limits_{-\pi/2}^{\pi/2} |\sin(\alpha_1 -\alpha_2)| f_{\alpha}(\alpha_1) f_{\alpha}(\alpha_2) \, \mathrm{d}\alpha_1\mathrm{d}\alpha_2.
\end{equation}
The quantity~\eqref{eq:sin12} is utilized to find the excluded area which is used to calculate the probability of the intersection of sticks and the average number of junctions (see Appendix~\ref{sec:deriv}). In the case of ODF~\eqref{eq:ODF4}, analytical solutions for~\eqref{eq:sin12} can be found only for some integer values of~$m$.
\begin{table}[!htbp]
  \centering
   \caption{Quantities of interest for different ODFs.}\label{tab:ODFs}
   \begin{tabular}{lccc}
     \hline
     \\
     ODF &  $\langle |\sin(\alpha_1 - \alpha_2)| \rangle$ & $\left \langle S^2 \right\rangle$ \\
     \\
     \hline
     \\
     \eqref{eq:ODF1} &  $\dfrac{2}{\pi}$ & $\dfrac{1}{2}$ \\
     \\
     \eqref{eq:ODF2} &  $ 1-\omega$ & $ 1$ \\
     \\
     \eqref{eq:ODF3} &  $\dfrac{1}{4\varepsilon^{2}} \left(2\varepsilon + 1 - \sin 2\varepsilon - \cos 2\varepsilon \right)$ & $ \dfrac{1}{2}+\dfrac{\sin 4\varepsilon}{8\varepsilon}$ \\
     \\
     \eqref{eq:ODF4}  & $\dfrac{13855460621}{7699613922 \pi}\approx 0.5727988619$\footnotemark[1] & $\dfrac{m+1}{m+2}$ \\
     \\
     \hline
   \end{tabular}
\end{table}
   \footnotetext[1]{$m=10$ }

\begin{figure}[!htbp]
  \centering
  \includegraphics[width=0.7\textwidth]{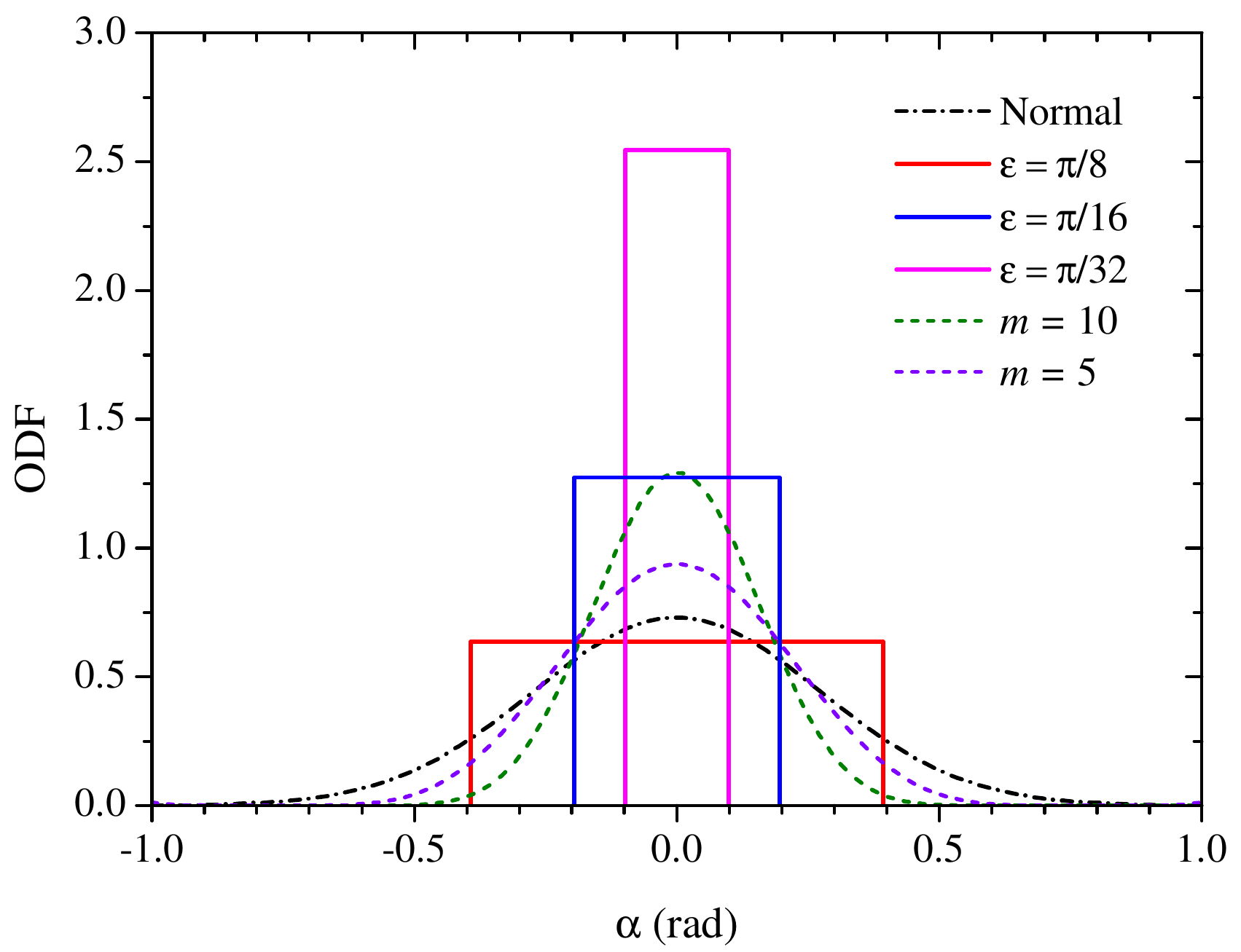}
  \caption{Different ODFs used in our study. Only one part of each distribution is shown. Another part of the same shape is located around the value $\pi/2$. Dash-dotted curve corresponds to~\eqref{eq:ODFGauss}. Solid lines correspond to~\eqref{eq:ODF3}. Dashed curve corresponds to~\eqref{eq:ODF4}. }\label{fig:ODFs}
\end{figure}

Figure~\ref{fig:ODFs} presents plots of ODFs used in our study. Although the distributions are not identical, significant impact of the particular ODF on the results is hardly expected.

\subsection{Simulation}\label{subsec:simul}

Conductive sticks were mimicked by linear segments of length $l=1$. These segments were placed within a square domain of linear size $L =32 $.  Periodic boundary conditions were applied to reduce the finite size effect. The centres of the segments were randomly placed within the domain. Their orientations of segments obeyed one of the desired ODFs. In such a way, a network was obtained, which served as a model of conducting films based on cross-aligned nanowires.

To detect the percolation cluster, the Union--Find algorithm~\cite{Newman2000PRL,Newman2001PRE} modified for continuous systems~\cite{Li2009PRE,Mertens2012PRE} was utilized. When a percolation  cluster was found, all other clusters were ignored since they cannot contribute to the electrical conductivity. An adjacency matrix was formed for the percolation cluster.  Having the adjacency matrix in hand, a random resistor network (RRN) was constructed.  Each contact between any two sticks was treated as a junction with an  electrical resistance $R_\text{j}$. A segment of stick between two nearest contacts with a separation $l_k$ corresponded to  a resistor with an electrical resistance, $R_\text{s} l_k/l$. In such a way, an RRN consisting of two kinds of resistors was considered.

Ohm’s law was applied to each resistor of any kind. Kirchhoff's current law was utilized for each junction between any two resistors. The resulting set of linear equations with a sparse matrix was solved using \emph{Eigen}~\cite{eigenweb}, a C++ template library for linear algebra.

The computer experiments were repeated 10 times for each value of the number density. The ratio of the stick resistance to the junction resistance were $\Delta = 10^6, 1, 10^{-6}$. The error bars in the figures demonstrating the dependence of the effective electrical conductivity on the number density are not shown since standard errors of mean are of the order of the marker size.

\begin{figure}[!htbp]
  \centering
  \includegraphics[width=0.7\textwidth]{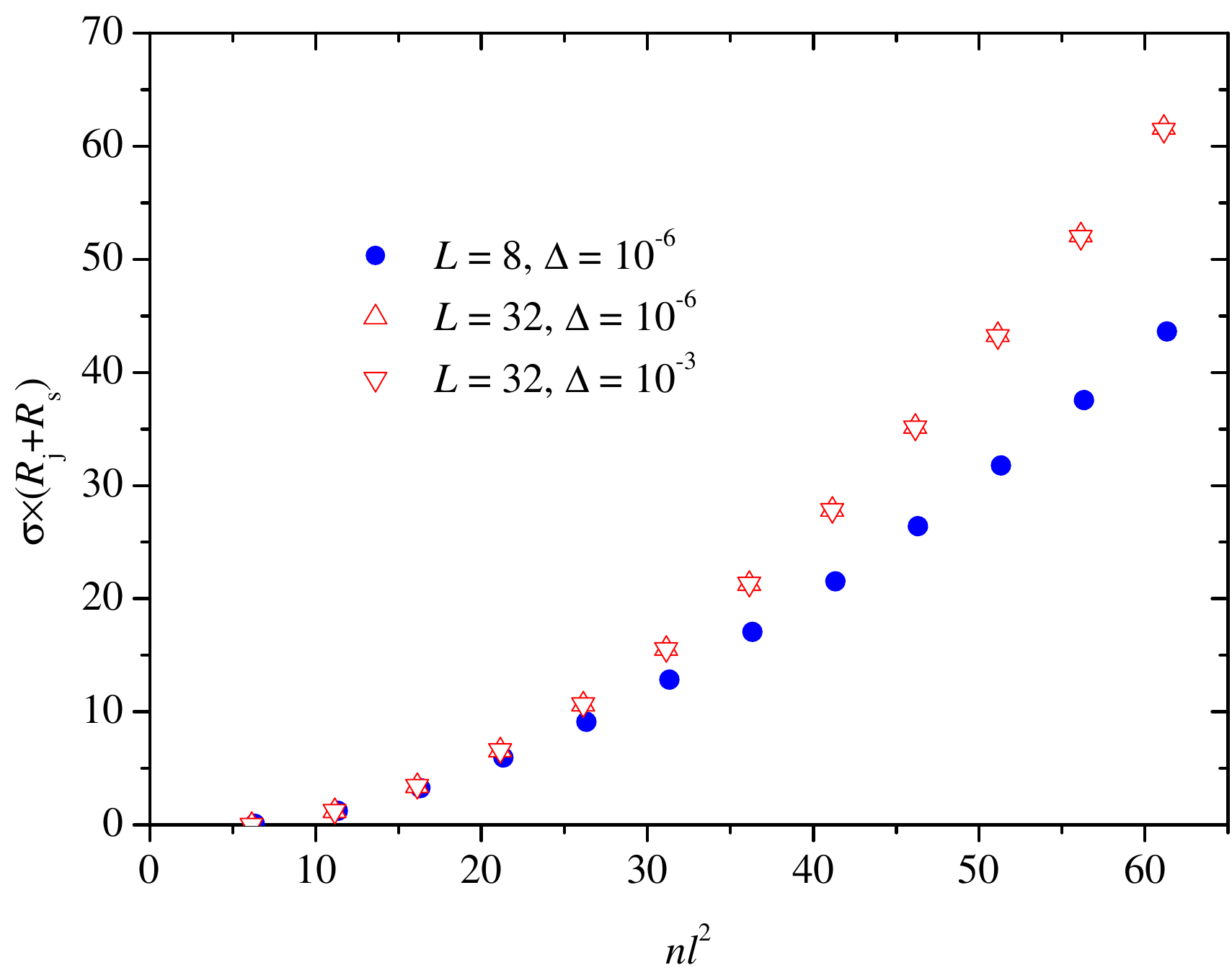}
  \caption{The dimensionless electrical conductivity vs the number density of conductive sticks for two different linear sizes of the system under consideration ($L=8$ and $L=32$ while $\Delta = 10^{-6}$) and for two different values of $\Delta$ ($\Delta = 10^{-6}$ and $\Delta = 10^{-3}$ while $L=32$). One half of the sticks is oriented along the $x$ axis, while the other half is oriented along the $y$ axis.}\label{fig:FSE}
\end{figure}

We performed test calculations to determine the finite size effect and effect of the value of the parameter $\Delta$. For  values of $\Delta = 10^{-6}$ and $\Delta = 10^{-3}$ ($L=32$), the electrical conductivities were consistent within the statistical error for all values of the number density (Fig.~\ref{fig:FSE}). In such a way, even $\Delta = 10^{-3}$ can be considered as a limiting case $\Delta \ll 1$. By contrast, despite the PBCs, the finite size effect is fairly visible, namely,  the electrical conductivity increases as the system size enlarges.

\section{Results}\label{sec:results}
\subsection{Analytical solution for the effective conductivity}\label{subsec:analytics}

Within the MFA (see details in Appendix~\ref{sec:deriv}), the effective electrical conductivity can be written as follows
\begin{equation}\label{eq:MFAsigma-RC1}
\sigma  \left(R_\text{s} + R_\text{j}\right)=  n l^2 \frac{1 + \Delta}{\Delta} \left\langle \cos^2 \alpha \right\rangle
\left[ 1-\frac{2}{\Delta}\sum_{N_\text{j}=2}^N f(N_\text{j},N,p)\frac{\left(\lambda_1-1\right)\left(\lambda_1^{{  N_\text{j}  }+1}-1\right)}{\lambda_1\left(\lambda_1^{  N_\text{j}  }+1\right)}\right].
\end{equation}
For convenience of further analysis, the effective electrical conductivity is written in a dimensionless form. Here, $N_j$ is the number of intersections of sticks with the given stick, $f(N_\text{j},N,p)$ is the fraction of sticks with exactly $N_\text{j}$ contacts. The number of intersections per stick is supposed to obey the binomial distribution $N_\text{j} \sim \Bin(N,p)$,
$$
f(N_\text{j},N,p) = \binom{N}{N_\text{j}}p^{N_\text{j}}(1-p)^{N-N_\text{j}},
$$
where $p$ is the probability that two arbitrary sticks intersect each other.
\begin{equation}\label{eq:Delta}
\Delta = \frac{R_\text{s}}{R_\text{j}},
\end{equation}
\begin{equation}\label{eq:C2}
\left\langle \cos^2 \alpha \right\rangle = \int\limits_{-\pi/2}^{\pi/2} f_{\alpha}(\alpha)
  \cos^2 \alpha \, \mathrm{d}\alpha,
\end{equation}
where $f_{\alpha}(\alpha)$ is the ODF;
\begin{equation}\label{eq:lambda1}
\lambda_1 = \frac{\mu + \sqrt{\mu^2 - 4}}{2},
\end{equation}
\begin{equation}\label{eq:mu}
  \mu = 2 + \frac{\Delta }{N_\text{j} + 1}.
\end{equation}

Derivation of the master equation~\eqref{eq:MFAsigma-RC1} presented in Appendix~\ref{sec:deriv} is rather rigorous. Therefore, Eq.~\eqref{eq:MFAsigma-RC1} can be considered accurate within the framework of the model under consideration. However, a further rigorous evaluation of the sum seems to be hardly possible. Since the MFA is applicable to dense systems only, we can expect a large number of intersections per stick on average.  The value of $N$ is supposed to be huge. This means that the binomial distribution tends to the Poisson distribution, which in turn tends to the Gaussian distribution and further to the Dirac $\delta$-function. Thus, in the case of dense systems, we can replace summation with integration with Dirac $\delta$-function.
\begin{equation}\label{eq:MFAsigma-RC0}
\sigma  \left(R_\text{s} + R_\text{j}\right)=  n l^2 \frac{1 + \Delta}{\Delta} \left\langle \cos^2 \alpha \right\rangle
\left[ 1-\frac{2}{\Delta}\frac{\left(\lambda_1-1\right)\left(\lambda_1^{{ \langle N_\text{j} \rangle }+1}-1\right)}{\lambda_1\left(\lambda_1^{ \langle N_\text{j}\rangle  }+1\right)}\right].
\end{equation}
Here,
\begin{equation}\label{eq:lambdaRC}
\langle N_\text{j} \rangle = n l^2 \langle |\sin(\alpha_1 - \alpha_2)| \rangle.
\end{equation}
When $ N_\text{j}  \to \infty$ and $\Delta$ is finite,
\begin{equation}\label{eq:lambda1asymp}
\lambda_1 \propto 1 + \sqrt{\frac{\Delta}{N_\text{j} }},
\end{equation}
\begin{equation}\label{eq:A}
A = \frac{\left(\lambda_1-1\right)\left(\lambda_1^{{  N_\text{j}  }+1}-1\right)}{\lambda_1\left(\lambda_1^{ N_\text{j} }+1\right)}
 \propto \lambda_1 -1 \propto \sqrt{\frac{\Delta}{N_\text{j} }}.
\end{equation}
The asymptotic estimate of the mathematical expectation is
\begin{equation}\label{eq:meanAasymp}
\mathbb{E}\left[A \right] \approx \sqrt{\frac{\Delta }{\langle N_\text{j} \rangle}} \left( 1 + \frac{3 }{8\langle N_\text{j} \rangle} \right),
\end{equation}
while the asymptotic behavior of the corresponding term in~\eqref{eq:MFAsigma-RC0} is
\begin{equation}\label{eq:termasymp}
\sqrt{\frac{\Delta }{\langle N_\text{j} \rangle}}.
\end{equation}
Hence, \eqref{eq:MFAsigma-RC0} systematically overestimates the electrical conductivity in the case of dense systems and finite values of $\Delta$. This behavior is expected to be also valid in other cases.

Thus, the improved formula for the asymptotic behavior of electrical conductivity can be written as follows:
\begin{equation}\label{eq:MFAsigma-RC0impr}
\sigma  \left(R_\text{s} + R_\text{j}\right)=  n l^2 \frac{1 + \Delta}{\Delta} \left\langle \cos^2 \alpha \right\rangle
\left[ 1-\frac{2}{\sqrt{\Delta  n l^2 \langle |\sin(\alpha_1 - \alpha_2)| \rangle} }\left( 1 +  \frac{ 3}{8 n l^2 \left\langle \left|\sin\left(\alpha_1 - \alpha_2\right)\right|\right\rangle} \right)\right].
\end{equation}
In our consideration, all conductors are assumed to participate in the electrical conductivity. Actually, there is no electrical conductivity below the percolation threshold, while only conductors belonging to the backbone of the percolation cluster participate in electrical conductivity above the percolation threshold. The effect of ODF on the percolation threshold was investigated in Ref.~\cite{Chatterjee2024}. Using the notation of our current article, the dependence of the percolation threshold on the cross-alignment can be written as follows
\begin{equation}\label{eq:PT}
  n_c l^2 = \frac{n_0}{\left\langle \left|\sin\left(\alpha_1 - \alpha_2\right)\right|\right\rangle},
\end{equation}
where $n_0$ is a constant. It is obvious that the cross-alignment of the sticks leads to a slight increase in the percolation threshold, which is quite consistent with the results of simulations~\cite{Grazioli2025}.

The current carrying part of the percolation threshold is called the backbone. According to Ref.~\cite{Kim2018JAP},  in the case of zero-width sticks, the backbone fraction can be written as
\begin{equation}\label{eq:BBKim2018}
  P_\text{bb} = 1 - \frac{2}{ N p} + \left( 1 + \frac{2(1-p) }{N p}\right)(1-p)^{N-1}.
\end{equation}
Similar equation has also been derived in Ref.~\cite{Kumar2017JAP}.
According to Ref.~\cite{Tarasevich2021PREbb}, for the ODF~\eqref{eq:ODF1} and large values of the nanowires, \eqref{eq:BBKim2018} can be rewritten as follows
\begin{equation}\label{eq:BBTarasevich2021}
  P_\text{bb} = 1 - \frac{\pi}{ n  l^2} + \left( 1 + \frac{\pi }{n  l^2}\right)\exp\left(-\frac{2 n l^2}{\pi}\right).
\end{equation}
The predictions of formula~\eqref{eq:BBTarasevich2021} are in excellent agreement with the results of computer modeling when $n \gtrapprox 2n_c$~\cite{Tarasevich2021PREbb}.

\begin{figure}[!htbp]
  \centering
  \includegraphics[width=0.7\textwidth]{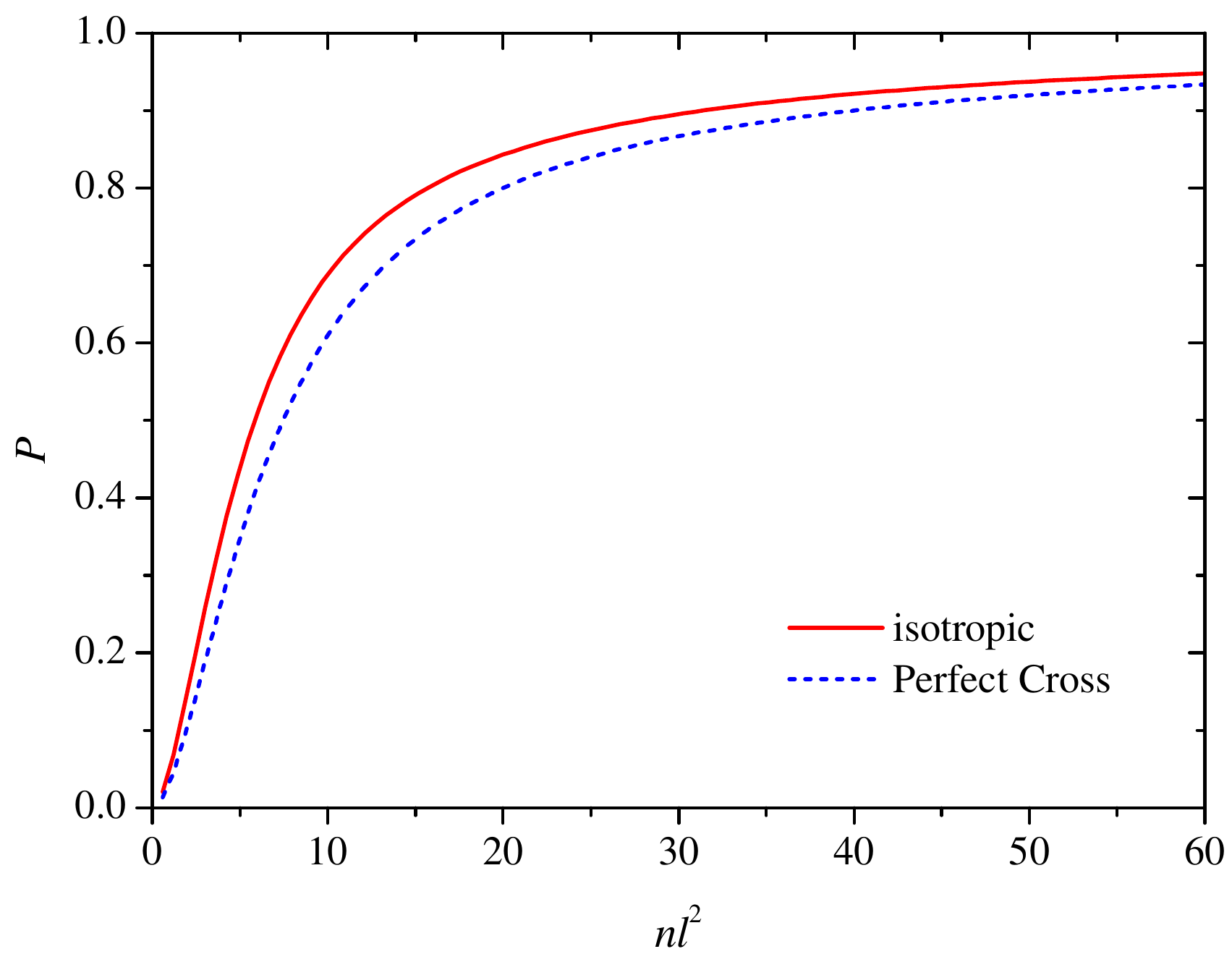}
  \caption{Examples of backbone fractions obtained using~\eqref{eq:BB} for the two limiting cases, namely, for the isotropic distribution~\eqref{eq:ODF1} and for the perfect cross~\eqref{eq:ODF3}.}\label{fig:BB}
\end{figure}

In the case of an arbitrary ODF and large values of the nanowires, Eq.~\eqref{eq:BBKim2018} can be written as follows
\begin{equation}\label{eq:BB}
    P_\text{bb} = 1 - \frac{2}{ n  l  ^2 \langle |\sin\left( \alpha_1 - \alpha_2 \right)|\rangle} + \left( 1 + \frac{2 }{n  l  ^2 \langle |\sin\left( \alpha_1 - \alpha_2 \right)|\rangle}\right)\exp\left(- n  l  ^2 \langle |\sin\left( \alpha_1 - \alpha_2 \right)|\rangle\right).
\end{equation}
Figure~\ref{fig:BB} demonstrates that the across-alignment of sticks slightly decreases the backbone fraction. Thus, at the same number density of nanowires, the fraction of conductors participating in electrical conductivity is higher in the case of a system with an equally probable orientation of nanowires than in the case of a system with cross-aligned nanowires. This means that, within the framework of a two-dimensional nanowire network model, orientational order leads to a decrease in electrical conductivity.

\subsection{Comparison of the analytical solution with simulations}\label{subsec:compar}
\begin{figure}[!htbp]
  \centering
  \includegraphics[width=0.7\textwidth]{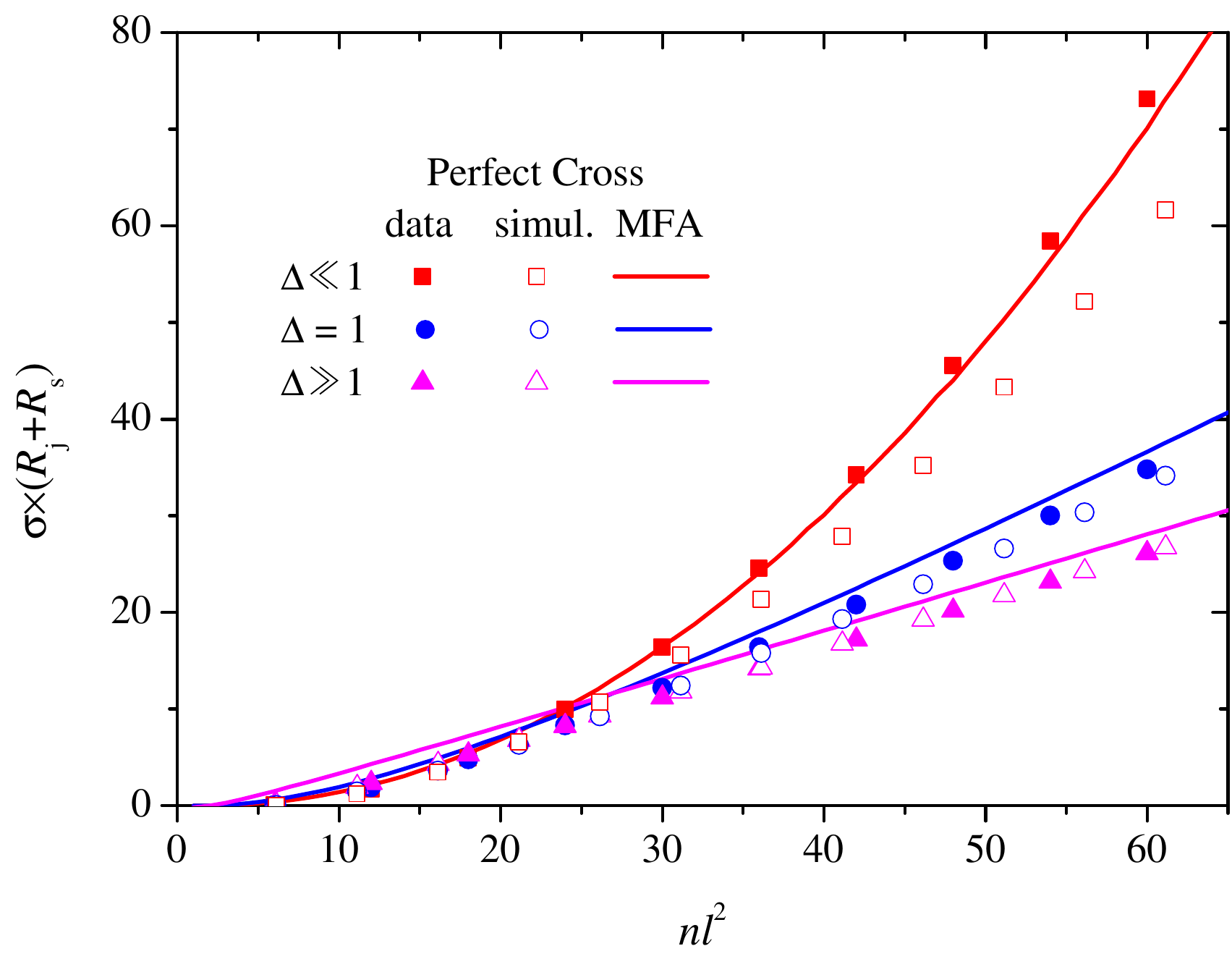}
  \caption{The dimensionless electrical conductivity of a system which obeys the ODF~\eqref{eq:ODF2} ($\omega=0.5$) vs the number density of conductive wires for the three different values of $\Delta$. Filled markers correspond to results published in~\cite{Grazioli2025}, open markers correspond to our direct computations, while curves present the MFA prediction~\eqref{eq:MFAsigma-RC0}.}\label{fig:PerfectCross}
\end{figure}

Figure~\ref{fig:PerfectCross} compares our direct computations of the electrical conductivity with  MFA estimate~\eqref{eq:MFAsigma-RC0} for the ODF~\eqref{eq:ODF2} ($\omega=0.5$). The electrical conductivity computed in Ref.~\cite{Grazioli2025} is also presented for comparison. When the junction resistance is negligible, our computations coincide with those presented in Ref.~\cite{Grazioli2025}. When the junction resistance and the stick resistance are equal, the results are close to each other. When the junction resistance dominates over the stick resistance, our results lie below the curve predicted by the MFA~\eqref{eq:MFAsigma-RC0}, in full agreement with our analysis of the accuracy of~\eqref{eq:MFAsigma-RC0}. By contrast, data extracted from  Ref.~\cite{Grazioli2025} exceed the prediction of the MFA, which is in conflict with our accuracy analysis. The MFA not only captures the trends correctly but also presents fairly close quantitative agreement with the direct computations. Note that the case $\Delta \gg 1$ is a limit that is never reached in real systems, and should therefore be considered only as a theoretical lower limit that is unattainable in practice. Contrary, the opposite case,  $\Delta \ll 1$, is more realistic. As a matter of fact, the dependencies for real-world systems are expected to lie between the curves for cases $\Delta = 1$ and $\Delta \ll 1$. Since the MFA knows nothing about the percolation threshold, it predicts an existence of the electrical conductivity up to zeroth number density, which is indeed wrong. Besides, the MFA supposes that the entire set of nanowires rather than the percolation cluster contributes the electrical conductivity. Although almost all nanowires belong to the percolation cluster at large values of the number density, the strength of the percolation cluster should be taken into account in the vicinity of the percolation threshold.

\begin{figure}[!htbp]
  \centering
  \includegraphics[width=0.7\textwidth]{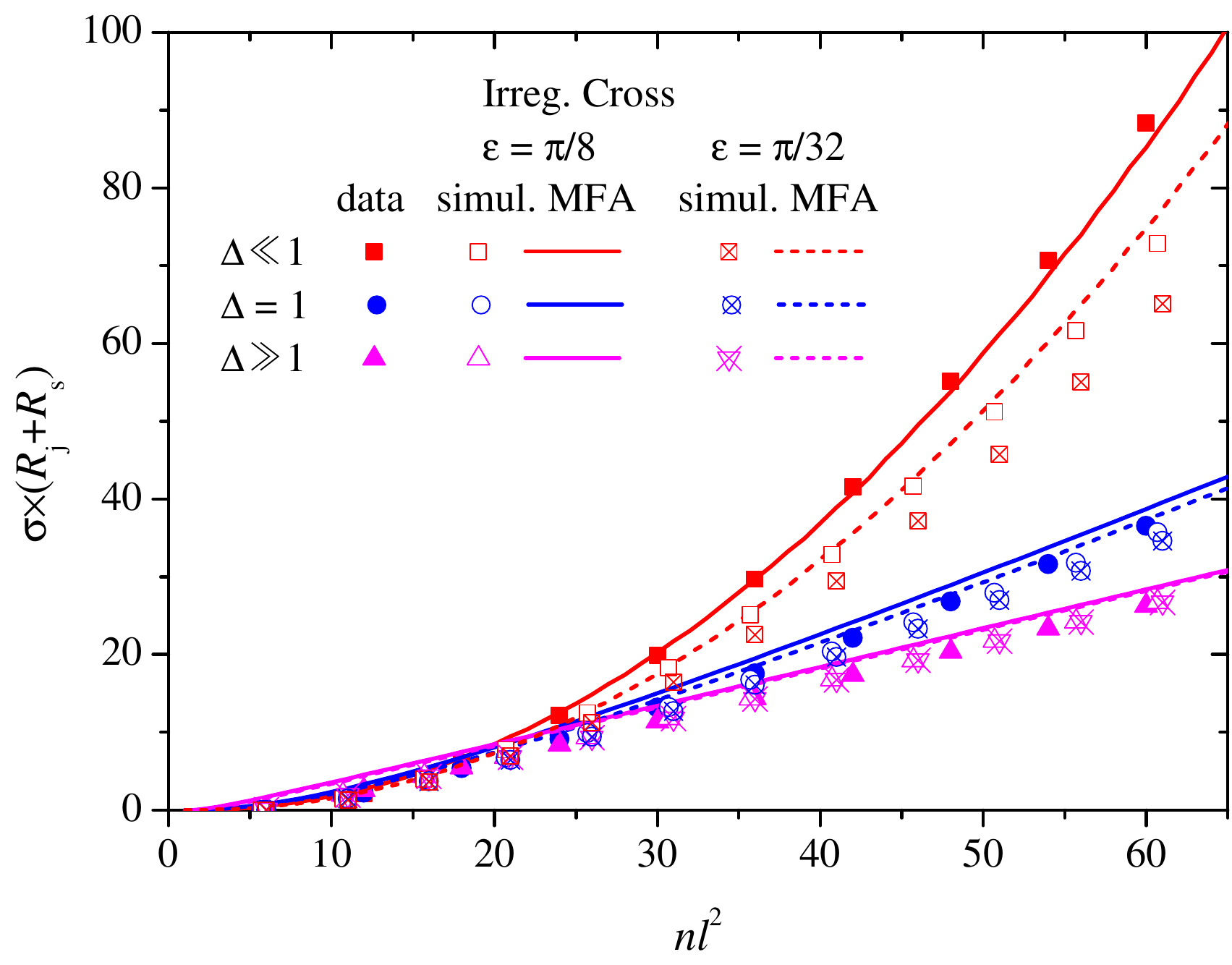}
  \caption{The dimensionless electrical conductivity of a system which obeys the ODF~\eqref{eq:ODF3}  vs the number density of conductive wires for the three different values of $\Delta$. $\left \langle S^2 \right\rangle = 0.818$ when $\varepsilon = \pi/8$, $\left \langle S^2 \right\rangle = 0.687$ when $\varepsilon = \pi/32$. Open markers correspond to the direct computations, while curves present the MFA prediction~\eqref{eq:MFAsigma-RC0}. For purpose of comparison, results published in~\cite{Grazioli2025} presented by filled markers.}\label{fig:IrregularCross}
\end{figure}

Figure~\ref{fig:IrregularCross} compares our direct computations of the electrical conductivity for the irregular cross with the predictions of the MFA using the ODF~\eqref{eq:ODF3}. Two values of the parameter $\varepsilon$ were used to produce irregular cross-aligned systems of nanowires. Thus, $\varepsilon = \pi/32$ corresponds to a narrow distribution, while $\varepsilon = \pi/8$ gets a wider distribution of nanowire orientations (see Fig.~\ref{fig:ODFs}). The electrical conductivity computed in Ref.~\cite{Grazioli2025} is also presented for comparison. Since the ODF~\eqref{eq:ODFGauss} was used in Ref.~\cite{Grazioli2025}, this can only be a comparison of trends, not a quantitative comparison. Figure~\ref{fig:IrregularCross} suggests that the width of the distribution has a small effect on the electrical conductivity, especially in the cases when the junction resistance is not a dominating factor. When the junction resistance is a dominating factor, the wider distribution the smaller the electrical conductivity.

\begin{figure}[!htbp]
  \centering
  \includegraphics[width=0.7\textwidth]{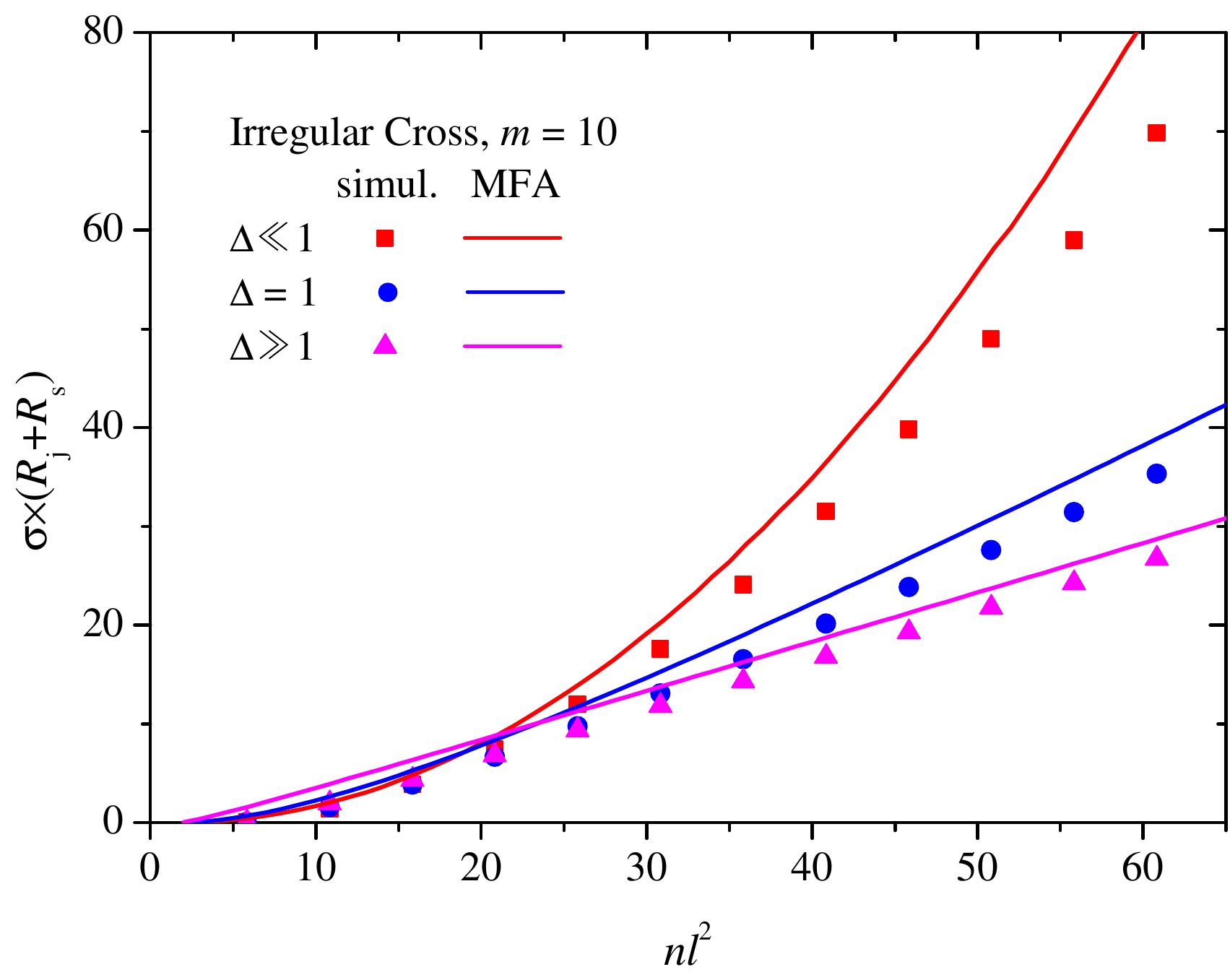}
  \caption{The dimensionless electrical conductivity of a system which obeys the ODF~\eqref{eq:ODF4} ($m=10$, $\left \langle S^2 \right\rangle = 0.917$)  vs the number density of conductive wires for the three different values of $\Delta$. Markers correspond to the direct computations, while curves present the MFA prediction~\eqref{eq:MFAsigma-RC0}. }\label{fig:ConductivityCosm10}
\end{figure}

Figure~\ref{fig:ConductivityCosm10} compares the direct computations of the electrical conductivity with the predictions of the MFA using the ODF~\eqref{eq:ODF4} with $m=10$.

Figures~\ref{fig:PerfectCross}, \ref{fig:IrregularCross}, and \ref{fig:ConductivityCosm10} show that, in full agreement with our estimate of the error that arises when replacing the mean value of a function with a function of the mean value of its argument~\eqref{eq:A}, \eqref{eq:meanAasymp}, and \eqref{eq:termasymp}, the MFA overestimates the value of electrical conductivity the more the smaller the value of the delta parameter.

\begin{figure}
  \centering
  \includegraphics[width=0.7\textwidth]{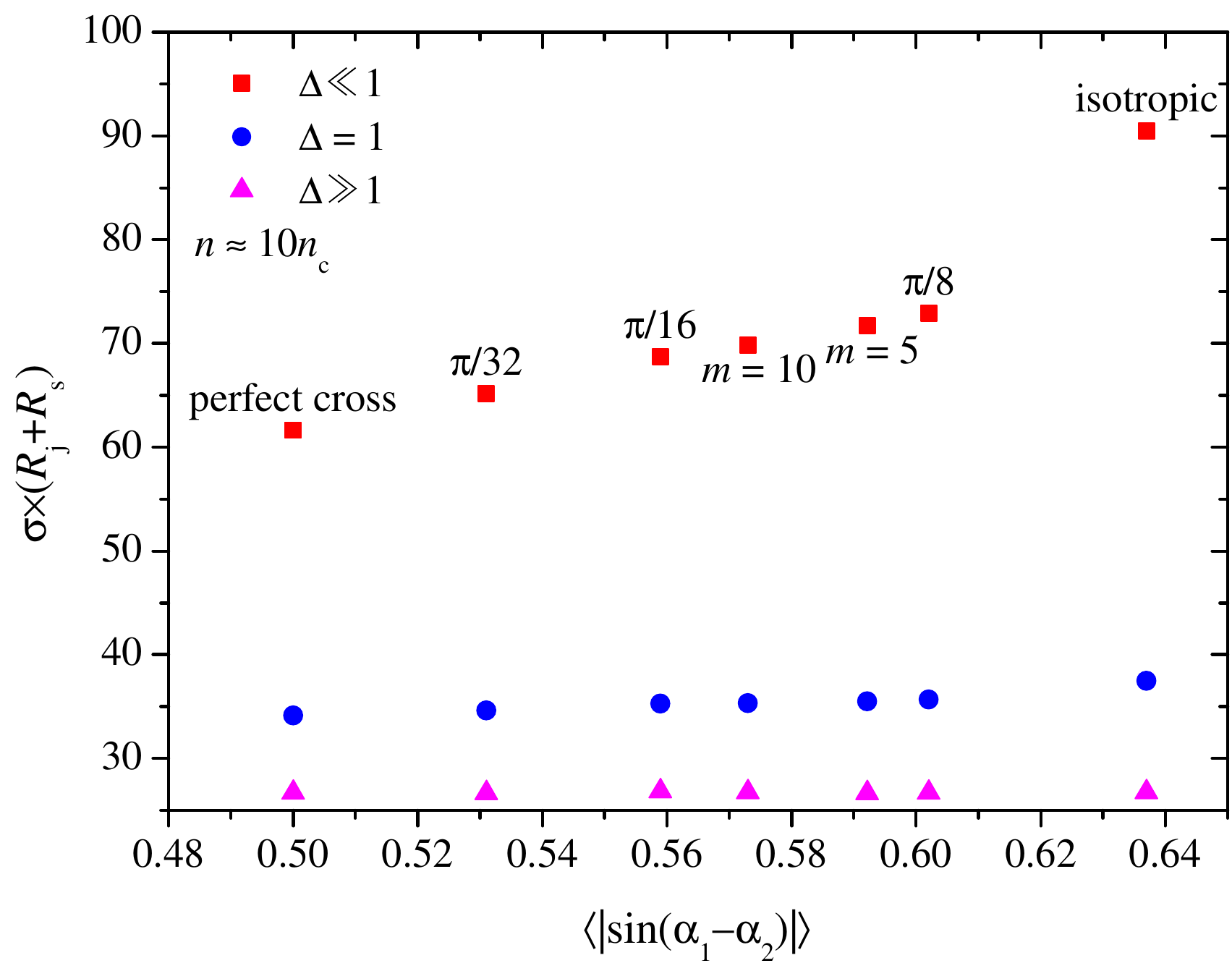}
  \caption{Dependence of electrical conductivity on ODFs for large values of the number density.}\label{fig:SigmavsSine}
\end{figure}

Figure~\ref{fig:SigmavsSine} suggests that the asymptotic behavior of the electrical conductivity  is insensitive to the particular ODF. The cross-alignment demonstrates no visible effect on the electrical conductivity when $\Delta \gg 1$. The effect of cross-alignment on the electrical conductivity is negligible  when $\Delta = 1$. The electrical conductivity slightly decreases from isotropic distribution to the cross-alignment when $\Delta \ll 1$.

\section{Conclusion}\label{sec:concl}
Using the mean-field approximation, we rigorously derived a formula for the effective electrical conductivity of a 2D system of randomly arranged conducting rods with a given orientation distribution. The formula takes into account both the resistance of the rods themselves and the resistance of the contacts between them. For practical application, the formula was simplified by replacing the mean value of the certain function with the function of the mean value of its argument. An accuracy estimate for this resulting formula demonstrated that such a replacement leads to a slight overestimate of the electrical conductivity. Theoretical predictions of the mean-field approximation were compared with the results of direct conductivity calculations for several model orientation distributions describing systems with crossed rods. The good agreement between the theoretical estimates and the direct calculation results demonstrates the feasibility of using the mean-field approximation to describe the electrical conductivity of conducting transparent films with crossed nanowires. Neither the calculations nor the mean-field theory predictions confirm the results of~\cite{Grazioli2025} for the case where junction resistance dominates.
Electrical conductivity does not depend on the specific type of the orientational distribution, but only on the average modulus of the sine of the angle between the conductors.
Cross-alignment reduces electrical conductivity when junction resistance dominates over the nanowire resistance and has no significant effect in other cases.
Cross-alignment leads to a slight increase in the percolation threshold, which is negative for the electrical conductivity.
Cross-alignment  slightly reduces the fraction of conductors belonging to the backbone of the percolation cluster, thus decreasing the electrical conductivity.
Summarizing, all considered effects of the nanowire cross-alignment have to reduce the electrical conductivity of transparent conductive films as compared to the films where the orientation of nanowires is equiprobable. Since experiments show the opposite, one can be concluded that the real cause lies outside the model. One of the possible explanations was proposed in Ref.~\cite{Grazioli2025}. The authors explain the observed effect due to decrease of the junction resistance in the case of cross-aligned nanowires. However, this explanation is not the only possible one. Although the ratio of the nanowires' length to their width is large, a zero-width stick is an extremely simplified model. Using zero-width sticks results in the actual quasi-three-dimensional (Q3D) network of nanowires being modeled using a 2D network with completely different properties~\cite{Daniels2021}. It was shown~\cite{Daniels2021} that in the 2D model, the average number of intersections per nanowire  increases linearly with increasing number density of nanowires, whereas in the Q3D model, the average number of intersections is almost independent of the number density of nanowires. In real-world systems of randomly distributed and equally probable oriented nanowires, the average number of contacts is expected to grow with increasing number density of nanowires  more slowly than linearly. Thus, the 2D model overestimates the connectivity of the nanowire network, leading to an overestimated electrical conductivity in both direct calculations and estimates within the MFA. On the other hand, in bilayer systems, such as those with cross-aligned nanowires, the 2D model is quite correct. However, the Q3D model also represents a limiting case, which is unlikely to be realized in the real world. In this model, nanowires are assumed to be infinitely rigid (hard rods of finite width), meaning they are incapable of any deformation. By contrast, using  zero-width sticks is somewhat equivalent to treating nanowires as infinitely soft, meaning they are capable of perfectly bending around each other (soft threads). Obviously, real nanowires exhibit some intermediate properties: they can be deformed, but this deformation does not ensure perfect bending.
We suppose that both 2D and Q3D networks are irrelevant models for isotropic nanowire networks.
In the case of an isotropic orientation distribution, the 2D model significantly overestimates the average number of contacts on a conductor compared to the Q3D model. This overestimation is particularly noticeable when junction resistance dominates over the wire resistance. However, the 2D network appears to be quite acceptable to model the two-layered transparent electrodes where all the nanowires in the first layer are aligned predominantly along one direction and in the perpendicular direction in the second layer.

Another possibility is that mutually repulsive forces may act when nanowires are aligned along the same direction. As a result, the arrangement of nanowire centers will differ from that obtained by modeling using a Poisson process to arrange the rods. In principle, this repulsion could easily be incorporated into computer modeling if appropriate experimental data were available, which we were unable to find.
Thus, the effect of nanowire repulsion during their placement on electrical conductivity of the final network remains an open question.

\section*{Acknowledgements}
Y.Y.T would like to thank Avik Chatterjee for a very valuable and fruitful discussion and  Davide Grazioli for data of their simulations.

\section*{Data availability}
The data that support the findings of this study are available from the corresponding author upon reasonable request.

\section*{Author contributions}
YYT: Supervision (lead), Conceptualization (lead), Methodology (lead), Visualization (lead), Writing – Original Draft Preparation (lead), Writing – Review \& Editing (lead);
AVE: Software (lead), Investigation (lead), Visualization (supporting);
IVV: Formal Analysis (lead), Writing – Review \& Editing (supporting).

\appendix
\section{Detailed derivation of the master equation}\label{sec:deriv}

\begin{figure}[!htb]
  \centering
  \includegraphics[width=0.7\columnwidth]{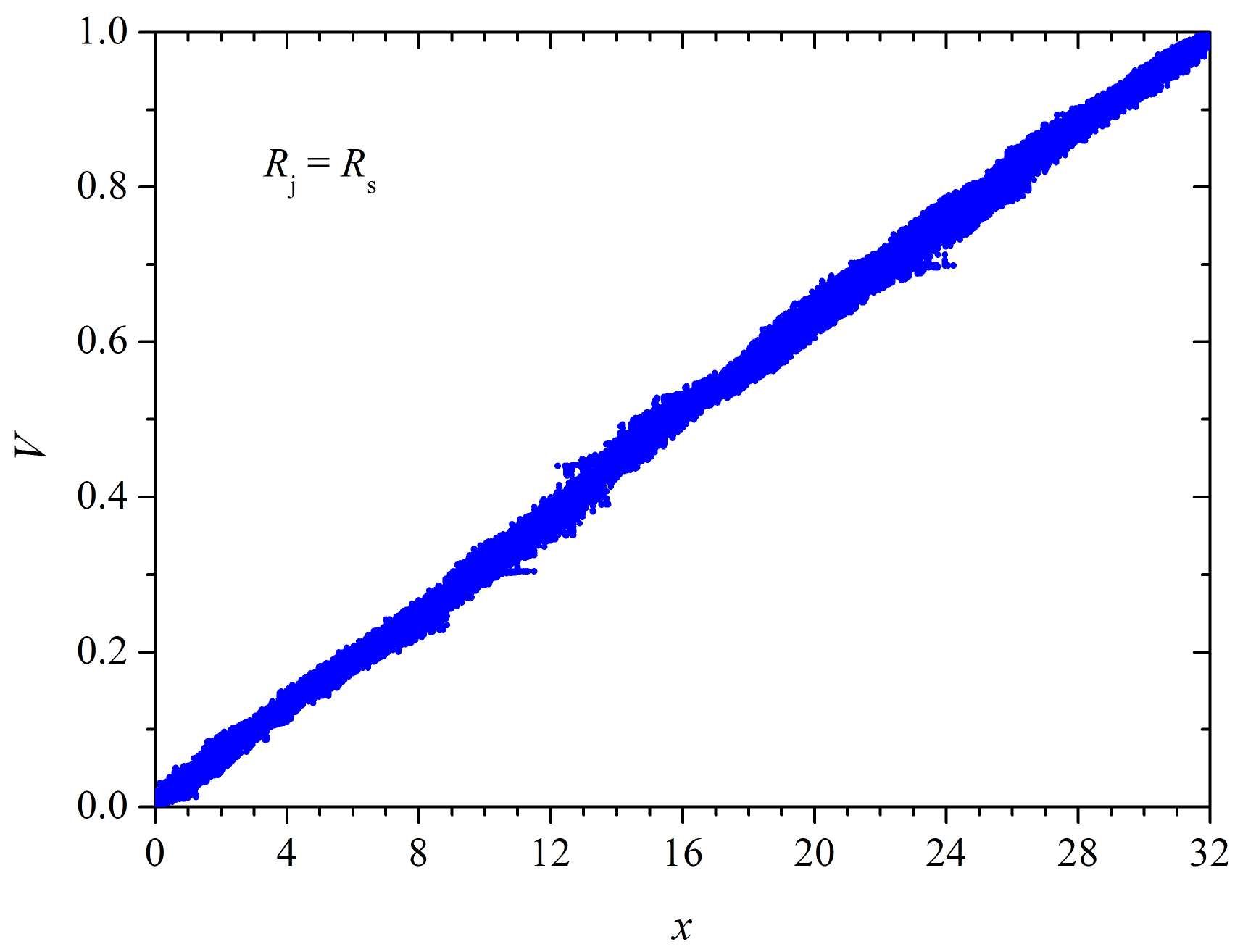}\\
  \caption{Example of simulated potential distribution in the system where all wires are divided into two  groups with equal numbers of sticks; in the first one, all wires are aligned along the $x$ axis, while in another, all wires are aligned along the $y$ axis. $L=32l$ when $n - n_c = 10$, where $n_c$ is the percolation threshold; for this particular system $n_c = 6.1934$, $R_\text{s}=R_\text{j}$.  \label{fig:potent}}
\end{figure}

Let a potential difference, $V_0$, be applied to the two opposite borders of the network under consideration. In dense systems, the voltage drop along the system is expected to be almost linear~\cite{Bergin2012Nanoscale,Khanarian2013JAP,Sannicolo2018,Forro2018ACSN,Papanastasiou2021,Charvin2021}.  This statement is also confirmed by our direct computations (Fig.~\ref{fig:potent}).  For seamless networks ($R_\text{j}=0$), the effect of the inhomogeneity of the electric field was taken into account~\cite{Tarasevich2023} and did not lead to any significant change in the main conclusions.

\begin{figure}[!htb]
  \centering
  \includegraphics[width=0.4\columnwidth]{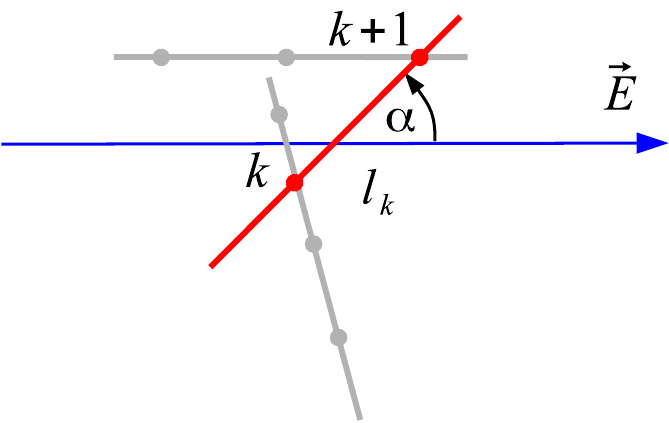}\\
  \caption{A linear conductor of length $l$ in an external electric field (red). Only two contacts are indicated. Their numbers are $k$ and $k+1$. The distance between these two contacts is $l_k$. The contact numbers go from 1 to $N_\text{j}$, while the numbers of segments, into which the contacts divide the conductor,  run from 0 to $N_\text{j}$.}\label{fig:contacts}
\end{figure}

Let there be a linear conductor of length $l$ and resistance $R_\text{s}$, which is immersed in an external homogeneous electric field $\mathbf{E}$. This field is directed along the $x$ axis. The angle between this conductor and the field is $\alpha$ (Fig.~\ref{fig:contacts}). Let $N_\text{j}$ point contacts with other conductors be distributed along this  conductor. These contacts provide some leakage conductance.

According to Ohm's law, the potential drop between the two nearest contacts $k$ and $k + 1$ equals
\begin{equation}\label{eq:u}
  u_{k+1} - u_k + \frac{l_k}{l} R_\text{s} i_k = 0,
\end{equation}
where $i_k$ is the electric current between the contacts $k$ and $k + 1$.

The change in the electric current when passing through contact number $k$ is due to a loss of charge in this contact
\begin{equation}\label{eq:i}
i_k - i_{k-1} + \frac{u_k - V_k}{R_\text{j}}  = 0,
\end{equation}
where $V_k$ is the potential of the external electrical field at the point where the contact of number $k$ is located.

Similarly,
\begin{equation}\label{eq:ik1}
i_{k+1} - i_{k} + \frac{u_{k+1} - V_{k+1}}{R_\text{j}}  = 0.
\end{equation}
Subtracting \eqref{eq:i} from \eqref{eq:ik1}, we obtain
\begin{equation}\label{eq:icomb}
i_{k+1} - 2 i_{k}  + i_{k-1} + \frac{1}{R_\text{j}}\left(u_{k+1} - u_k - V_{k+1} + V_k \right)= 0.
\end{equation}
Equations~\eqref{eq:u} and~\eqref{eq:icomb} can be combined
\begin{equation}\label{eq:inhomoi01}
i_{k+1} - 2 i_{k}  + i_{k-1} + \frac{1}{R_\text{j}}\left(-\frac{l_k}{l} R_\text{s} i_k - V_{k+1} + V_k \right)= 0
\end{equation}
or
\begin{equation}\label{eq:inhomoi1}
i_{k+1} - \left(2 +\Delta\frac{l_k}{l}  \right)i_{k}  + i_{k-1} = \frac{V_{k+1} - V_k }{R_\text{j}},
\end{equation}
where
\begin{equation}\label{eq:muDelta}
\Delta = \frac{R_\text{s}}{R_\text{j}}.
\end{equation}
Accounting for
\begin{equation}\label{eq:pot}
  V_{k+1} - V_k = -\frac{V_0 \cos \alpha}{L }l_k,
\end{equation}
one gets
\begin{equation}\label{eq:inhomoi001}
i_{k+1} - \left(2 +\Delta\frac{l_k}{l}  \right)i_{k}  + i_{k-1} = -\frac{V_0 \cos \alpha}{L R_\text{j}}l_k ,
\end{equation}

Let the ODF of the conductor segment lengths be $f_l(l_{k};N_\text{j})$. Multiply equation \eqref{eq:inhomoi001} by $f_l(l_{k};N_\text{j})$ and integrate from 0 to $l$, then
\begin{equation}\label{eq:inhomoi}
i_{k+1} - \left(2 +\frac{\Delta}{N_\text{j} + 1}  \right)i_{k}  + i_{k-1} = -\frac{V_0 \cos \alpha}{L R_\text{j}}\frac{l}{N_\text{j} + 1},
\end{equation}
since
\begin{equation}\label{eq:lk}
\int\limits_0^l l_k f_l(l_k;N_\text{j})\, \mathrm{d}l_k = \frac{l}{N_\text{j} + 1}.
\end{equation}
We have a non-homogeneous recurrence relation (difference equation) with constant coefficients
\begin{equation}\label{eq:udropmeanmu}
i_{k+1} - \mu i_{k} + i_{k - 1}  = - \frac{V_0 \cos \alpha}{L R_\text{j}}\frac{l}{N_\text{j} + 1}.
\end{equation}
Since the choice of the number $k$ is arbitrary, we can write a similar equation with the number shifted by 1.
\begin{equation}\label{eq:udropmeanmu1}
i_{k+2} - \mu i_{k+1} + i_{k}  = - \frac{V_0 \cos \alpha}{L R_\text{j}}\frac{l}{N_\text{j} + 1}.
\end{equation}
Subtracting \eqref{eq:udropmeanmu} from \eqref{eq:udropmeanmu1}, we obtain a homogeneous third-order recurrence relation with constant coefficients
\begin{equation}\label{eq:udropmeanmu2}
 i_{k+2} - (\mu + 1) i_{k+1} + (\mu + 1) i_{k} - i_{k - 1} = 0.
\end{equation}
We are looking for a solution in the form $i_k = \lambda^k$. Then
\begin{equation}\label{eq:charact}
\lambda^{k+2} - (\mu + 1) \lambda^{k+1} + (\mu + 1) \lambda^{k} - \lambda^{k - 1} = 0.
\end{equation}
Since the trivial solution $\lambda = 0$ is out of our interest, then, dividing by $\lambda^{k - 1}$, we have
\begin{equation}\label{eq:charact1}
\lambda^{3} - (\mu + 1) \lambda^{2} + (\mu + 1) \lambda - 1 = 0.
\end{equation}
Obviously, $\lambda_3 =1$, then we can find the remaining roots using the quadratic equation
\begin{equation}\label{eq:charact2}
\lambda^{2} - \mu \lambda + 1 = 0.
\end{equation}
\begin{equation}\label{eq:lambda23}
\lambda_{1,2} = \frac{\mu \pm \sqrt{\mu^2 - 4}}{2}.
\end{equation}
Note, that
\begin{equation}\label{eq:lambdaplus}
\lambda_{1} +\lambda_{2} = \mu,
\end{equation}
\begin{equation}\label{eq:lambdaprod}
 \lambda_{1} \lambda_{2} = 1.
\end{equation}
Therefore
\begin{equation}\label{eq:ikcommon}
i_k = a_0 + a_1 \lambda_1^k +a_2 \lambda_2^k.
\end{equation}
Since the terminal segments of any stick are dead ends, that is, electric current cannot flow through them, therefore $i_0 =0$ and $i_{N_\text{j}}= 0$, then
\begin{equation}\label{eq:a1a2a3begin}
a_0 + a_1 + a_2 =0
\end{equation}
and
\begin{equation}\label{eq:a1a2a3end}
a_0 + a_1 \lambda_1^{N_\text{j}} + a_2 \lambda_2^{N_\text{j}} = 0.
\end{equation}
From the equation~\eqref{eq:udropmeanmu} it follows that
\begin{equation}\label{eq:udropmeanmulambd}
a_0 + a_1 \lambda_1^{k+1} + a_2 \lambda_2^{k+1} - \mu \left( a_0 + a_1 \lambda_1^{k } + a_2 \lambda_2^{k }\right) + a_0 + a_1 \lambda_1^{k - 1} + a_2 \lambda_2^{k - 1}  = - \frac{V_0 l \cos \alpha}{\left(N_\text{j} + 1\right)L R_\text{j}}.
\end{equation}
Rearranging, we get
\begin{equation}\label{eq:udropmeanmulambd1}
a_0(2 - \mu ) + a_1 \lambda_1^{k - 1}\left( \lambda_1^2 - \mu \lambda_1 +1\right) + a_2 \lambda_2^{k - 1}\left( \lambda_2^{2} - \mu  \lambda_2 + 1 \right) = - \frac{V_0 l \cos \alpha}{\left(N_\text{j} + 1\right)L R_\text{j}}.
\end{equation}
Since $\lambda_1$ and $\lambda_2$ are the roots of the square equation~\eqref{eq:charact2}, then $\lambda_1^2 - \mu \lambda_1 +1 = 0$ and $\lambda_2^{2} - \mu  \lambda_2 + 1 =0$.
Accounting for~\eqref{eq:mu},
$$
2 - \mu  =  - \frac{\Delta}{N_\text{j} + 1}.
$$
Hence,
\begin{equation}\label{eq:a1}
  a_0 =  \frac{V_0  l\cos \alpha}{L R_\mathrm{s} }.
\end{equation}
From equations~\eqref{eq:a1a2a3begin} and~\eqref{eq:a1a2a3end}, accounting for~\eqref{eq:lambdaprod}, it follows that
$$
a_2 =  a_1 \lambda_1^{N_\text{j}}.
$$
Hence,
$$
a_1 =
-\frac{a_0}{1+ \lambda_1^{N_\text{j}}}.
$$
$$
a_2 =  - a_0 \frac{ \lambda_1^{N_\text{j}}}{1 + \lambda_1^{N_\text{j}}}.
$$

The solution of the linear recurrence with constant coefficients~\eqref{eq:charact1}
\begin{equation}\label{eq:isolution}
  i_k(\alpha,N_\text{j}) = \frac{V_0 l \cos \alpha}{L R_\text{s}} \left( 1 - \frac{ \lambda_1^k  +  \lambda_1 ^{ N_\text{j}} \lambda_1^{-k} }{ 1 + \lambda_1^{N_\text{j}}}\right)
\end{equation}
describes the electric current in the $k$-th segment of the linear conductor having exactly $N_\text{j}$ contacts, and which is located at an angle $\alpha$ to the external electric field.

The average electric current in a conductor with $N_\text{j}$  contacts and which is located at an angle $\alpha$ to the external electric field is equal to
\begin{equation}\label{eq:imeanalphaNjnew}
  \langle i(\alpha, N_\text{j})\rangle = \frac{V_0 l \cos \alpha}{L R_\text{s}}\left( 1 - \frac{1}{N_\text{j}+1}\sum_{k=0}^{N_\text{j}}\frac{ \lambda_1^k  +  \lambda_1 ^{ N_\text{j}} \lambda_1^{-k} }{ 1 + \lambda_1^{N_\text{j}}}\right).
\end{equation}
The sum can be easily calculated, which leads to
\begin{equation}\label{eq:imeanalphaNj1new}
  \langle i\left(\alpha,N_\text{j}\right)\rangle = \frac{V_0 l \cos \alpha}{L R_\text{s}}
  \left[ 1 - \frac{2 \left( \lambda_1^{ N_\text{j}  +1}  - 1 \right)}{\left( \lambda_1^{ N_\text{j}}  + 1 \right)\left( N_\text{j}  +1 \right)\left( \lambda_1  - 1 \right)}\right].
\end{equation}
Accounting for~\eqref{eq:mu}, \eqref{eq:lambdaplus}, \eqref{eq:lambdaprod},
$$
 N_\text{j}  +1  =  \frac{\Delta}{\mu -2} = \frac{\Delta \lambda_1}{\left( \lambda_1 - 1 \right)^2}.
$$
\begin{equation}\label{eq:imeanalphaNj1new1}
  \langle i\left(\alpha,N_\text{j}\right)\rangle = \frac{V_0 l \cos \alpha}{L R_\text{s}}
  \left[ 1 - \frac{2 }{\Delta }\frac{\left( \lambda_1 - 1 \right) \left( \lambda_1^{ N_\text{j}  +1}  - 1 \right)}{\lambda_1\left( \lambda_1^{ N_\text{j}}  + 1 \right)}\right].
\end{equation}
When $N_\text{j}=0$ (a solitary stick), \eqref{eq:imeanalphaNj1new1} yields 0, which is correct. When $N_\text{j}=1$ (only one intersection, i.e., both termini of the stick are dead-ends), \eqref{eq:imeanalphaNj1new1} gives again 0, which is true.

Then, the number of sticks with exactly $N_\text{j}$ contacts intersecting a line of length $L$ perpendicular to the electrostatic field is $ f(N_\text{j},N,p) n l L \cos \alpha $. The total electric current in the such sticks of all orientations through a cross section of the system is
\begin{multline}\label{eq:int}
f(N_\text{j},N,p) n l L  \int\limits_{-\pi/2}^{\pi/2} f_{\alpha}(\alpha) \left\langle i\left(\alpha, N_\text{j}\right)\right\rangle  \cos \alpha \, \mathrm{d}\alpha \\=
f(N_\text{j},N,p) \frac{n  l^2 V_0 }{ R_\text{s}}\left[ 1-\frac{2}{\Delta}\frac{\left(\lambda_1-1\right)\left(\lambda_1^{{  N_\text{j}  }+1}-1\right)}{\lambda_1\left(\lambda_1^{  N_\text{j}  }+1\right)}\right] \left\langle \cos^2 \alpha \right\rangle ,
\end{multline}
where
$$
\left\langle \cos^2 \alpha \right\rangle = \int\limits_{-\pi/2}^{\pi/2} f_{\alpha}(\alpha)
  \cos^2 \alpha \, \mathrm{d}\alpha
$$
and $f_{\alpha}(\alpha)$ is the appropriate ODF.

The total electric current  through a cross section of the system is
\begin{equation}\label{eq:itotalnew}
i=
 \frac{n  l^2 V_0 }{ R_\text{s}}\left\langle \cos^2 \alpha \right\rangle \sum_{N_\text{j}=2}^N f(N_\text{j},N,p)\left[ 1-\frac{2}{\Delta}\frac{\left(\lambda_1-1\right)\left(\lambda_1^{{  N_\text{j}  }+1}-1\right)}{\lambda_1\left(\lambda_1^{  N_\text{j}  }+1\right)}\right],
\end{equation}
According to Ohm's law, the electrical conductivity is
\begin{equation}\label{eq:MFAsigma-discr}
\sigma =
 \frac{n  l^2 }{ R_\text{s}}\left\langle \cos^2 \alpha \right\rangle \sum_{N_\text{j}=2}^N f(N_\text{j},N,p)\left[ 1-\frac{2}{\Delta}\frac{\left(\lambda_1-1\right)\left(\lambda_1^{{  N_\text{j}  }+1}-1\right)}{\lambda_1\left(\lambda_1^{  N_\text{j}  }+1\right)}\right],
\end{equation}
\begin{equation}\label{eq:MFAsigma-discr1}
\sigma =
 \frac{n  l^2 }{ R_\text{s}} \left\langle \cos^2 \alpha \right\rangle \left[ 1-\frac{2}{\Delta}\sum_{N_\text{j}=2}^N f(N_\text{j},N,p)\frac{\left(\lambda_1-1\right)\left(\lambda_1^{{  N_\text{j}  }+1}-1\right)}{\lambda_1\left(\lambda_1^{  N_\text{j}  }+1\right)}\right].
\end{equation}

In the limiting case when the sticks are superconductive, $R_\text{s}=0$, the above evaluations should be slightly modified. Eq.~\eqref{eq:u} transforms into
\begin{equation}\label{eq:usc}
  u_{k+1} - u_k  = 0,
\end{equation}
while \eqref{eq:inhomoi01} reads now as
\begin{equation}\label{eq:inhomoiSC}
i_{k+1} - 2 i_{k}  + i_{k-1} + \frac{1}{R_\text{j}}\left(- V_{k+1} + V_k \right)= 0.
\end{equation}
Since now coefficient $\mu=2$, Eq.~\eqref{eq:udropmeanmu2} is
\begin{equation}\label{eq:udropmeanmuSC}
 i_{k+2} - 3 i_{k+1} + 3 i_{k} - i_{k - 1} = 0.
\end{equation}
Its characteristic equation is
\begin{equation}\label{eq:charactSC}
\lambda^{3} - 3 \lambda^{2} + 3 \lambda - 1 = (\lambda - 1)^3 = 0,
\end{equation}
hence, $\lambda_1=\lambda_2=\lambda_3=1$.
$$
i_k = C_1 + k C_2 + k^2 C_3.
$$
$$
i_0 = C_1 =0.
$$
$$
i_{N_\text{j}} = N_\text{j} C_2 + N_\text{j}^2 C_3 = 0,
$$
hence,
$$
C_3 = - \frac{C_2 }{ N_\text{j}} .
$$
$$
i_k = k C_2 \left( 1 - \frac{k}{N_\text{j}}\right).
$$
Accounting for~\eqref{eq:inhomoi} with $\Delta=0$, the solution of this linear recurrence with constant coefficients is
\begin{equation}\label{eq:ikNj}
  i_k(\alpha,N_\text{j}) = k ( N_\text{j}  - k )\frac{V_0 l \cos \alpha}{2 L (N_\text{j} + 1) R_\text{j}}.
\end{equation}
The electric current in the conductor, averaged over all segments, is
\begin{equation}\label{eq:electriccurent}
\left\langle i(\alpha,N_\text{j}) \right\rangle = \frac{1}{N_\text{j} + 1}\sum_{k=0}^{N_\text{j}} i_k =
\frac{V_0 l \cos \alpha}{2 L (N_\text{j} + 1)^2 R_\text{j}}\sum_{k=0}^{N_\text{j}} k ( N_\text{j}  - k ).
\end{equation}
Calculating the sum leads to the formula
\begin{equation}\label{eq:iaver1}
\langle i(\alpha,N_\text{j}) \rangle =\frac{N_\text{j} (N_\text{j}-1)}{12 L R_\text{j}(N_\text{j} + 1)} V_0 l \cos \alpha.
\end{equation}
The total electric current in all sticks with exactly $N_\text{j}$ contacts of all orientations through a cross section of the system is
\begin{equation}\label{eq:totali}
\langle i(N_\text{j}) \rangle
= C(N_\text{j}) nl^2 \frac{N_\text{j} (N_\text{j}-1)}{12 L R_\text{j}(N_\text{j} + 1)}V_0 l \left\langle \cos^2 \alpha \right\rangle.
\end{equation}

The number of contacts on the conductor obeys the binomial distribution which, as the number of sticks is large, tends to Poisson distribution~\cite{Heitz2004NT,Yi2004JAP,Callaghan2016PCCP,Kumar2017JAP,Kim2018JAP}. Thus, the average current in the conductor is
\begin{equation}\label{eq:iaver2}
\langle i \rangle = \frac{V_0 l }{12 L R_\text{j}} \sum_{k=0}^{\infty}\frac{k(k-1) \left( n l^2 \left\langle \cos^2 \alpha \right\rangle \right)^k }{(k + 1) k!} \mathrm{e}^{-\left\langle \cos^2 \alpha \right\rangle n l^2} = \frac{V_0 l }{12 L R_\text{j}} D(n),
\end{equation}
where
\begin{equation}\label{eq:Dvsn}
D(n) = \left\langle \cos^2 \alpha \right\rangle n l^2 - 2   - 2\frac {1 - {\rm e}^{-\left\langle \cos^2 \alpha \right\rangle n l^2} }{\left\langle \cos^2 \alpha \right\rangle n l^2}.
\end{equation}

Note, $D(n) \approx  n l^2 \left\langle \cos^2 \alpha \right\rangle$, when $ n l^2 \left\langle \cos^2 \alpha \right\rangle \gg 1$ .

The number of sticks intersecting a line of length $L$ perpendicular to the field is $ n l L \cos \alpha $.
Hence, the electrical conductivity of the system under consideration is
\begin{equation}\label{eq:MFAsigmaJDRdiscr}
  \sigma =  \frac{ n l^2 }{24 R_\text{j}} D(n).
\end{equation}
When $ n l^2 \left\langle \cos^2 \alpha \right\rangle \gg 1$ , then
\begin{equation}\label{eq:MFAsigmaJDRcont}
\sigma = \frac{\left(n l^2 \right)^2 }{12 \pi R_\text{j}}.
\end{equation}

\begin{figure}[!htbp]
  \centering
  \includegraphics[width=0.3\textwidth]{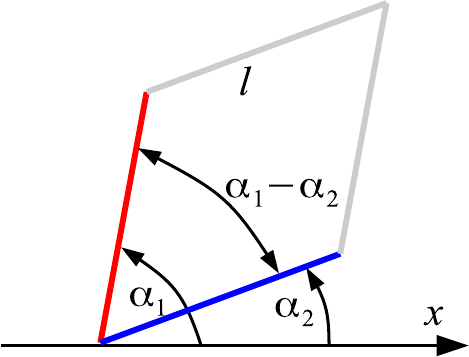}
  \caption{The excluded area of the blue rod is shown as a rhombus.}\label{fig:romb}
\end{figure}

According to the binomial distribution, the mean number of intersections per stick is
$$
\langle N_\text{j} \rangle = p N,
$$
where $p$ is the probability that two arbitrary sticks intersect each other.
If the angle between the two sticks is $\alpha_1 -\alpha_2$, then the probability of their intersection is equal to the ratio of the excluded area, $A_\text{ex}=l^2\left|\sin\left(\alpha_1 -\alpha_2\right)\right|$, to the area of the entire domain, $L^2$,
$$
p(\alpha_1 -\alpha_2) = \frac{l^2}{L^2} \left|\sin\left(\alpha_1 -\alpha_2\right)\right|,
$$
see, e.g., \cite{Kim2018JAP} (Fig.~\ref{fig:romb}).

The probability of intersection of two arbitrary sticks is
$$
\int\limits_{-\pi/2}^{\pi/2} p(\alpha_1 -\alpha_2)f_{\alpha}(\alpha_1) f_{\alpha}(\alpha_2) \, \mathrm{d}\alpha_1\mathrm{d}\alpha_2 = \frac{l^2}{L^2} \int\limits_{-\pi/2}^{\pi/2} \left|\sin\left(\alpha_1 -\alpha_2\right)\right| f_{\alpha}(\alpha_1) f_{\alpha}(\alpha_2) \, \mathrm{d}\alpha_1\mathrm{d}\alpha_2.
$$
Therefore,
$$
\langle N_\text{j} \rangle = n l^2 \int\limits_{-\pi/2}^{\pi/2} \left|\sin\left(\alpha_1 -\alpha_2\right)\right| f_{\alpha}(\alpha_1) f_{\alpha}(\alpha_2) \, \mathrm{d}\alpha_1\mathrm{d}\alpha_2.
$$

For the ODF~\eqref{eq:ODF1}, the probability of intersection is
$$
p = \frac{ l^2}{\pi L^2} \int\limits_{0}^{\pi/2} \sin\alpha \, \mathrm{d}\alpha = \frac{2 l^2}{\pi L^2},
$$
hence,
$$
\langle N_\text{j}\rangle  = \frac{2 n l^2}{\pi }.
$$
For the ODF~\eqref{eq:ODF2}, the probability of intersection is
$$
p = (1-\omega)\frac{ l^2}{L^2}  ,
$$
hence,
$$
\langle N_\text{j} \rangle = (1-\omega) n l^2.
$$
When $\omega=1/2$,
$$
\langle N_\text{j} \rangle = \frac{ n l^2}{2}.
$$

For~\eqref{eq:ODF3}, $z = \alpha_1-\alpha_2$
$$
f_Z(z) = \begin{cases}
\dfrac{2\varepsilon - |z|}{8\varepsilon^2}, & |z| \leq 2\varepsilon, \\
\\
\dfrac{2\varepsilon - \left|z - \frac{\pi}{2}\right|}{16\varepsilon^2}, & \frac{\pi}{2} - 2\varepsilon \leq z \leq \frac{\pi}{2} + 2\varepsilon, \\
\\
\dfrac{2\varepsilon - \left|z + \frac{\pi}{2}\right|}{16\varepsilon^2}, & -\frac{\pi}{2} - 2\varepsilon \leq z \leq -\frac{\pi}{2} + 2\varepsilon, \\
\\
0, & \text{otherwise}.
\end{cases}
$$
$$
\mathbb{E}[|\sin(z)|] = \frac{1}{4\varepsilon^{2}} \left(2\varepsilon + 1 - \sin 2\varepsilon - \cos 2\varepsilon \right),
$$
$$
\langle N_\text{j} \rangle = \frac{n l^2}{4\varepsilon^{2}} \left(2\varepsilon + 1 - \sin 2\varepsilon - \cos 2\varepsilon \right).
$$

\bibliography{CrossAlignedNWs}
\bibliographystyle{unsrt}

\end{document}